# AN APPLICATION OF PRINCIPAL STRATIFICATION TO CONTROL FOR INSTITUTIONALIZATION AT FOLLOW-UP IN STUDIES OF SUBSTANCE ABUSE TREATMENT PROGRAMS[1]

By Beth Ann Griffin, Daniel F. McCaffrey
and Andrew R. Morral

*RAND Corporation*

Participants in longitudinal studies on the effects of drug treatment and criminal justice system interventions are at high risk for institutionalization (e.g., spending time in an environment where their freedom to use drugs, commit crimes, or engage in risky behavior may be circumscribed). Methods used for estimating treatment effects in the presence of institutionalization during follow-up can be highly sensitive to assumptions that are unlikely to be met in applications and thus likely to yield misleading inferences. In this paper we consider the use of principal stratification to control for institutionalization at follow-up. Principal stratification has been suggested for similar problems where outcomes are unobservable for samples of study participants because of dropout, death or other forms of censoring. The method identifies principal strata within which causal effects are well defined and potentially estimable. We extend the method of principal stratification to model institutionalization at follow-up and estimate the effect of residential substance abuse treatment versus outpatient services in a large scale study of adolescent substance abuse treatment programs. Additionally, we discuss practical issues in applying the principal stratification model to data. We show via simulation studies that the model can only recover true effects provided the data meet strenuous demands and that there must be caution taken when implementing principal stratification as a technique to control for post-treatment confounders such as institutionalization.

**1. Introduction.** Each year almost 1.8 million Americans receive alcohol and other drug treatment services. Efforts to improve these services through

Received June 2007; revised April 2008.

[1]Supported by NIDA Grants R01 DA015697, R01 DA016722 and R01 DA017507.

research or provider profiling are hindered by drug treatment clients' high rates of post-treatment institutionalization, defined here as spending a day or more in a controlled environment (e.g., a jail, prison, hospital, residential treatment or group home setting) where the possibility of drug use and criminal activity is substantially diminished. By reducing the potential for substance use, institutionalization masks the potential effects of substance use treatment programs. For example, outcomes of clients from both effective and ineffective treatment programs will look the same when they have no access to drugs or alcohol.

The confounding of treatment effects due to institutionalization cannot be ignored when evaluating substance abuse treatment programs, because institutionalization is so pervasive among drug treatment clients. For instance, in the Drug Abuse Treatment Outcomes Study [Hubbard et al. (1997)], 40% of the 2,966 clients in substance abuse treatment programs interviewed 12-months after discharge reported being institutionalized for some part of the preceding year [U.S. Dept. of Health and Human Services, National Institute on Drug Abuse (2004)]. Among those with any institutionalization, the average number of days institutionalized out of the past 365 was 115 [U.S. Dept. of Health and Human Services, National Institute on Drug Abuse (2004)]. Similarly, about 52% of the sample from the National Treatment Improvement Evaluation (NTIES) [U.S. Dept. of Health and Human Services, Substance Abuse and Mental Health Services Administration, Center for Substance Abuse Treatment (2004), NORC and RTI (1997)] were institutionalized during the study's 12-month post-treatment evaluation, many for the entire length of the evaluation period.

The confounding effects of institutionalization are not, however, limited to substance abuse treatment programs. Criminal justice system clients are also at great risk for being institutionalized following an intervention. Moreover, patients involved in many health studies are at risk for hospitalizations which can limit or suppress the measurement of primary outcomes of interest in those studies (e.g., daily physical activities for a general population or falls for geriatric patients). In all of these cases, a post-treatment factor (e.g., institutionalization or hospitalization) whose value is determined after the start of treatment consequently determines the potential range of the outcome that can be observed, thereby suppressing or censoring the outcome of primary interest. In many situations, inferences are further complicated by the fact that the confounding factor can take on many levels giving rise to outcomes which are observed at different values of the confounding variable and invalidating the comparability of outcomes in the treatment and control groups.

In a recent paper McCaffrey et al. (2007) developed a statistical model to describe the different possible estimates of the causal effect of treatment in the presence of institutionalization and the potential confounding effects of institutionalization on these estimates. The paper identifies commonly used analytic methods for estimating treatment effects in the presence of institutionalization and demonstrates that the estimated treatment effects can vary greatly depending on which method is employed. The paper also identifies assumptions under which the various approaches yield unbiased estimates of the treatment effects of interest. Unfortunately many of the assumptions required appear unlikely to hold in most real world applications.

Institutionalization is similar to the problem of incomplete compliance in drug trials. In both problems, cases assigned to treatment have multiple potential outcomes that can vary depending on a variable not controlled by the experimenter: dose of the drug (i.e., level of compliance) in drug trials or days institutionalized in our example. Hence, methods for estimating the dose-response curve from partial compliance data [Efron and Feldman (1991)] might apply to the institutionalization problem. However, these methods make substantial use of the fact that treatment is a complete dose of the drug, whereas, in our example institutionalization is not related to the amount or type of treatment received but is an external event that curtails use censoring the outcomes of interest.

The problem of institutionalization is more similar to the problems addressed by the methods of principal stratification. Developed by Frangakis and Rubin (2002) for problems where outcomes are unobservable for samples of study participants because of dropout, death or other forms of censoring (e.g., failure to find employment in a jobs program), the method identifies principal strata within which causal effects are well defined and potentially estimable. Roughly speaking, Frangakis and Rubin (2002) noted that in data with censoring, cases that receive treatment and have observed outcomes are a mix of cases that would have observed outcomes under control and those that would not. Similarly, cases that receive the control condition and have observed outcomes are a mix of cases that would have observed outcomes under treatment and those that would not. A study participant's censoring statuses under the treatment and control conditions define the strata within the population, called principal strata. Analyses that condition on cases in the stratum where participants would be observed under both treatment and control can provide unbiased treatment effect estimates for that stratum assuming no other biasing factors. The key innovation to the work of principal stratification is the recognition that conditioning on censoring statuses under treatment and control is sufficient to allow for unbiased estimation of treatment effects.

Principal strata notions extend to the problem of institutionalization at follow-up with only slight modifications to the methods for censored data. In particular, modifications are necessary because institutionalization can take on more than the two levels of censored and uncensored. Also, outcomes are observed at all levels of institutionalization and treatment effects at various levels may be of interest.

Using data from the Adolescent Treatment Models (ATM) study, fielded between 1998 and 2002 by the Substance Abuse and Mental Health Services Administration, Center for Substance Abuse Treatment, we aim to examine the effects of treatment modality (residential versus outpatient) on the 12-month substance use outcomes for adolescents who participated in the ATM using principal stratification. In the ATM, high rates of institutionalization clearly confound the effects of treatment modality. Adolescents in residential treatment have significantly higher rates of institutionalization and lower mean drug use frequency outcomes at the 12-month follow-up. Given that institutionalization in the sample can be shown to yield lower drug use outcomes [McCaffrey et al. (2007)], it is unclear whether adolescents in residential treatment truly have lower mean values of the outcome because the treatment is more effective than outpatient treatment or because they tend to be institutionalized more often.

The goal of this paper is to extend principal stratification to control for the confounding effects of institutionalization and to evaluate the sensitivity of the extended model to various assumptions about the data. Section 2 describes the data from the ATM study and illustrates the confounding effects institutionalization is likely to have when examining treatment effects in this study. Section 3 describes the method of principal stratification and its extension in more detail and applies the method to the ATM data to examine the effects of substance use treatment modality (residential versus outpatient) on drug use outcomes for adolescents in the study. Section 4 evaluates the principal stratification method presented using a series of simulation studies and Section 5 provides a discussion of our findings and recommendations for practice and future research.

**2. Adolescent treatment models study and the institutionalization confound.**  The number of adolescents receiving substance abuse treatment has increased by over 65% in the last 10 or so years and policy makers, clinicians and parents want to know if treatment is effective and for whom it is effective, as well as what treatment modalities are best. We address these questions through a study of the effects of treatment modality (residential versus outpatient) on the 12-month substance use outcomes for adolescents who participated in the ATM study. The ATM study collected treatment admission and 12-month outcomes data for new admissions to 10 community-based treatment programs in the United States, including six residential programs and four outpatient programs [Stevens and Morral (2003)].

The sample used in the present analysis includes all new admissions to the 10 programs in the main ATM analytic dataset, which was produced in March of 2002. Of these 1,384 cases, 1,256 (91%) completed a 12-month follow-up survey and provided data on the outcomes of interest. Only cases with follow-up data are included in the analysis presented below.

For purposes of the illustrations presented in this report, we examine the effects of treatment modality on the Substance Frequency Scale (SFS), a widely used scale from the Global Appraisal of Individual Needs (GAIN) [Dennis (1999)], the survey instrument used at every site for baseline and 12-month outcome assessments. The SFS averages responses to a series of questions on the frequency of recent drug use, intoxication and drug problems in the 90 days prior to the 12 month follow-up.[1] It is scaled so that higher values indicate greater substance use and more drug problems. Days of institutionalization at follow-up is assessed with the maximum number of days—in the past 90—which the respondent reports being in any of several different types of controlled environments (e.g., inpatient psychiatric or medical hospitals, residential treatment facilities, juvenile halls or other criminal justice detention facilities, etc.).

Figure 1 shows a scatter plot and smooth of mean SFS versus the number of months institutionalized for adolescents enrolled in residential and outpatient care treatment modalities in the ATM. As shown, more adolescents in residential treatment have months of institutionalization greater than 0. In fact, 52% of adolescents in residential care were institutionalized for at least one day in the past 90 at the 12-month follow-up, while only 38% of adolescents in outpatient care were institutionalized at follow-up (see Table 1). Figure 1 also reveals the suppression effect that institutionalization can have on SFS. As the number of months institutionalized increases, the observed values of SFS in the ATM data appear to decrease. This relationship can be seen more clearly in the smooth of SFS versus days institutionalized also shown in the figure.

In addition to the suppression effect of institutionalization, Figure 1 also clearly reveals the potential selection effects that exists in this data between adolescents with and without any days of institutionalization. The mean value of SFS for adolescents who are not institutionalized at follow-up (represented by the black X above 0 days institutionalized in Figure 1) is markedly lower than the mean values for adolescents with only a few days of institutionalization (shown along the smoothed line in Figure 1). The difference in mean SFS between these adolescents suggests that adolescents who were institutionalized in the 90 days prior to the 12 month follow-up tended to be more difficult cases with higher levels of drug use than those adolescents not entering institutional settings.

Both the suppression and selection effects of institutionalization in the ATM data could distort inferences about the effects of treatment on SFS.

---

[1]In this illustration we multiplied SFS by 90 to make it scale with use in the past 90 days rather than use per day.

Table 1

*Weighted mean rate of institutionalization and SFS at 12-month follow-up by treatment modality*

|                      | **Percent institutionalized at follow-up** | **Mean SFS** |
|----------------------|:---:|:---:|
| Residential treatment | 52% | 9.0 |
| Outpatient treatment  | 38% | 9.7 |

Table 1 on page 6 provides weighted descriptive statistics (see below for more details on weighting) for SFS and institutionalization at follow-up by treatment modality. As shown, adolescents in residential treatment have higher rates of institutionalization and lower mean SFS. Given that institutionalization appears to suppress substance use in this sample, it is unclear whether adolescents in residential treatment have lower mean values of SFS because they have decreased their substance use in response to residential treatment or because they tend to be institutionalized more often.

Unfortunately, as described in McCaffrey et al. (2007), current methods for handling institutionalization are not adequate and require strong assumptions which are unlikely to hold in practice. In light of these findings, we propose the use of principal stratification to obtain policy-relevant treat-ment effects on this data which appropriately control for the suppression and selection effects institutionalization can have on SFS.

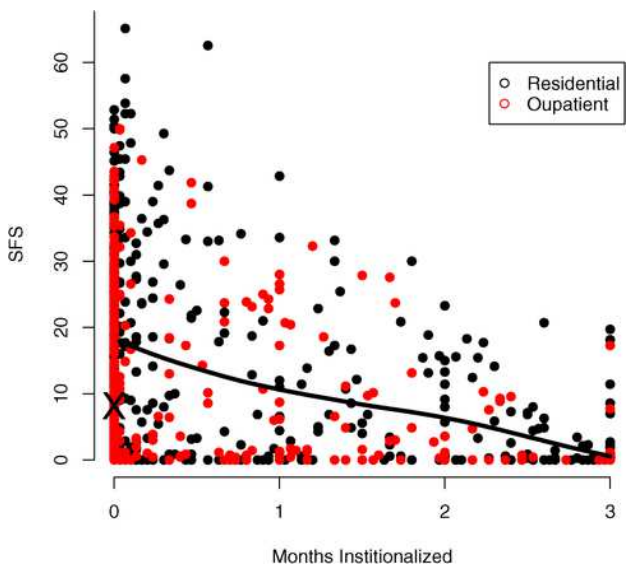

Fig. 1. *Scatter plot of SFS values by number of months institutionalized for adolescents in both residential and outpatient treatment modalities with smooth of mean SFS by months institutionalized overlayed. The black* X *denotes the mean SFS value among adolescents with 0 days of institutionalization.*

**3. Principal stratification.** Principal stratification was developed by Frangakis and Rubin ([2002](#)) as a method for accounting for post-treatment confounds within the context of the Neyman–Rubin causal model. Hence, we begin by extending the Neyman–Rubin causal model to account for institutionalization and then turn to describe the specific innovations of the principal stratification approach.

3.1. *A causal model for treatment effects in the presence of institutionalization.* We start by considering the treatment effect of a single intervention versus a control in the simple case without any institutionalization. In this case, the Neyman–Rubin causal model [Holland ([1986](#)), Pearl ([1996](#))] considers two potential outcomes and one random variable for each case in the study. The potential outcomes are $Y_0$, the outcome after receiving the control condition, and $Y_1$, the outcome after receiving the treatment condition. Throughout we assume that the Stable Unit Treatment Value Assumption (SUTVA) [Rubin ([1990](#))] holds so that for each case the potential outcomes are unique and well defined.

The treatment effect for each individual is $Y_1 - Y_0$. Typically this is summarized by its mean across study participants. However, we cannot directly estimate the treatment effect for individual cases or the mean across cases because cases cannot be observed under both the treatment and the control conditions. The condition under which each case is observed is determined by the random treatment assignment variable $T$, which equals 1 for treatment and 0 for control. When $T = 1$, we observe $Y_1$, otherwise $Y_0$ is observed. This results in the random variable $Y_{obs} = Y_T$.

Unbiased estimation of the average treatment effect is possible if cases with $T = 1$ have the same expected values of their potential outcomes as cases with $T = 0$. Under this assumption, $E(Y_{obs}|T = 1) = E(Y_1)$ and $E(Y_{obs}|T = 0) = E(Y_0)$, where expectation is over the participating cases. Hence, the difference in the observed treatment and control means yields an unbiased estimate of the average causal effect of treatment.

Consistent estimation is also possible in situations where treatment assignment might depend on observable characteristics, such as observational studies where study participants self-select into the treatment programs being studied. In such circumstances, consistent estimates can be obtained by comparing the means for cases with the same probability of treatment and using propensity score weights to adjust for selection effects into a particular treatment [Rosenbaum and Rubin ([1983](#))].

In the presence of institutionalization, the Neyman–Rubin causal model described above must be expanded to allow for cases to have potential levels of institutionalization, which we denote by $Z_0$ and $Z_1$ for the control and treatment conditions, respectively. The model must also allow for different potential outcomes at each level of institutionalization. For example, a case might have a different potential outcome when institutionalized 0 days, compared to 1, 2 or 3 or more days. We let $Z_{\max}$ equal the maximum possible value for institutionalization. Then, for treatment, $T = 1$, we label the potential outcomes for a case as $Y_1[z]$, $z = 0, \ldots, Z_{\max}$ so that $Y_1[0]$ is the potential outcome if assigned to treatment and not institutionalized during the follow-up period and $Y_1[1]$ is the potential outcome if assigned to treatment and had 1 day of institutionalization, and so on to $Z_{\max}$. The potential outcomes for control are $Y_0[z]$, $z = 0, \ldots, Z_{\max}$.

Now, for each case, we can define different treatment effects for each of the different levels of institutionalization, for example, $D[z] = Y_1[z] - Y_0[z]$ such that $D[z]$ might change with $z$. While there are multiple causal effects that might be of interest, not all may be estimable from the data without strong assumptions. We might, for example, be interested in the average causal effect of $D[0]$, the average causal effect had no one been institutionalized, which McCaffrey et al. (2007) refer to as the "unsuppressed treatment effect." Additionally, we might also want to estimate the average causal effect for cases at each of the observed values of institutionalization as was considered when the notion of principal stratification was derived [Frangakis and Rubin (2002)]. We now turn to that concept.

3.2. *Principal stratification.* To estimate the causal effects of treatment using the notions of principal stratification, we need a model for both the outcomes and institutionalization. Specifically, we assume that institutionalization can take on only a discrete set of values $0, 1, \ldots, Z_{\max}$ for both treatment and control. For example, if the follow-up interval for observing outcomes is 90 days, then $Z_1$ and $Z_0$ can only take on a value from 0 to 90.

Under this assumption, we can in turn define $Z_{\max} * Z_{\max}$ principal strata as shown in Table 2. We define the probability that an individual falls into stratum $(z_0, z_1)$ by $p_{z_0, z_1} = Pr\{Z_0 = z_0, Z_1 = z_1\}$ and define a density function for the potential outcome $Y_t[z_t] | Z_0 = z_0, Z_1 = z_1$ if a person falls into this stratum by $f(y; \boldsymbol{\theta}_{z_0, z_1, t})$ for $z_0, z_1 = 1, \ldots, Z_{\max}$, where $\boldsymbol{\theta}_{z_0, z_1, t}$ denotes the parameter vector for the assumed underlying density function $f$ and $t$ denotes whether or not an individual received treatment ($t = 1$) or control ($t = 0$). We note that the probabilities, $p_{z_0, z_1}$, do not depend on treatment status because $Z_0$ and $Z_1$ are potential outcomes and a case's principal stratum remains the same regardless of which treatment he/she receives. Conversely, the distribution of the potential outcomes, $f$, is determined by the principal strata and treatment status of an individual. For example, if we can assume the potential outcomes are normally distributed, then we might have separate means and variances in $\boldsymbol{\theta}_{z_0, z_1, t}$ for each possible pair of values of $(z_0, z_1)$ for both treatment and control.

TABLE 2
*Principal stratification model for institutionalization under treatment $t = 0, 1$*

|  |  | $Z_0 = 0$ | 1 | · · · | $Z_{\max}$ |
|---|---|---|---|---|---|
| $Z_1 =$ | 0 | $(p_{00}, f(y; \boldsymbol{\theta}_{0,0,t}))$ | $(p_{10}, f(y; \boldsymbol{\theta}_{1,0,t}))$ | · · · | $(p_{Z_{\max},0},$ $f(y; \boldsymbol{\theta}_{Z_{\max},0,t}))$ |
|  | 1 | $(p_{01}, f(y; \boldsymbol{\theta}_{0,1,t}))$ | $(p_{11}, f(y; \boldsymbol{\theta}_{1,1,t}))$ | · · · | $(p_{Z_{\max},1},$ $f(y; \boldsymbol{\theta}_{Z_{\max},1,t}))$ |
|  | 2 | $(p_{02}, f(y; \boldsymbol{\theta}_{0,2,t}))$ | $(p_{12}, f(y; \boldsymbol{\theta}_{1,2,t}))$ | · · · | $(p_{Z_{\max},2},$ $f(y; \boldsymbol{\theta}_{Z_{\max},2,t}))$ |
|  | . | . | . | · · · | . |
|  | . | . | . | · · · | . |
|  | . | . | . | · · · | . |
|  | $Z_{\max}$ | $(p_{0,Z_{\max}},$ $f(y; \boldsymbol{\theta}_{0,Z_{\max},t}))$ | $(p_{1,Z_{\max}},$ $f(y; \boldsymbol{\theta}_{1,Z_{\max},t}))$ | · · · | $(p_{Z_{\max},Z_{\max}},$ $f(y; \boldsymbol{\theta}_{Z_{\max},Z_{\max},t}))$ |

The primary treatment effects of interest in this model are $E[D[z]|Z_0 = z, Z_1 = z]$, that is, the treatment effects for individuals with the same level of institutionalization under both treatment and control. These treatment effects do not confound changes in institutionalization with other effects of treatment and allow policy-makers, stake holders and caregivers to evaluate treatment independent of the potentially costly and undesirable effects on institutionalization. Each average effect is restricted to cases from a principal stratum. Generalizations to other strata require additional assumptions.

We let $S_{z_t,t}$ denote the set of all cases in condition $t$ whose observed value of institutionalization equals $z_t$. Then if $y_i$ denotes the observed outcome for the $i$th case, the likelihood for the observed data is given by

$$L(\mathbf{Y}, \mathbf{Z}, \mathbf{T}, \boldsymbol{\theta}) = \prod_{z_1} \prod_{i \in S_{z_1,1}} \sum_{z_0} f(y_i; \boldsymbol{\theta}_{z_0, z_1, 1}) p_{z_0, z_1}$$

$$(3.1)$$

$$\times \prod_{z_0} \prod_{i \in S_{z_0,0}} \sum_{z_1} f(y_i; \boldsymbol{\theta}_{z_0, z_1, 0}) p_{z_0, z_1}.$$

This mixture distribution results from the fact that only $Z_1$ or $Z_0$ is observed for each case. A similar likelihood was presented for an application of principal stratification methods for estimating the causal effect of a jobs program [Zhang, Rubin and Mealli (2005)]. In their model, $Z_{\max}$ was 1 and the potential outcomes $Y_t$ did not exist when $Z_t = 1$. In our case, outcomes can exist at every level of institutionalization.

Weights can be easily incorporated into the likelihood function in equation (3.1), allowing for maximization of a weighted likelihood function. A natural application of such a weighted likelihood includes the maximization of the likelihood when comparing propensity score weighted clients in one treatment program to unweighted clients in another, as will be illustrated in Section 3.3 below.

For many common distributions, mixture likelihoods like (3.1) can be optimized using the EM algorithm. Theoretically, we can use the EM algorithm to estimate causal effects with any set of data. However, in practice, the solution is unlikely to be so simple. First, if there are many levels of $Z$ to model, there will be many possible levels of institutionalization; then, without any additional structure, there will be a very large number of parameters to estimate. To diminish the dimensionality of the problem, one can consider modeling $\boldsymbol{\theta}_{z_0, z_1, t}$ as low order polynomials in $z_0$ and $z_1$. For example, we might assume $E(Y_1 | Z_1 = z_1, Z_0 = z_0) = \mu + \beta_1 z_1 + \beta_0 z_0 + \gamma z_1 z_0$. Second, it is well known that the convergence of likelihood optimizers for mixture models can be sensitive to the starting values [Biernacki, Celeux and Gerard (2003), Karlis and Xekalaki (2003), McLachlan (1988), Seidel and Sevcikova (2004)]. Given the large numbers of mixtures involved in this likelihood and the fact that identification of the parameters will depend on matching the mixing proportions/probabilities across groups, sensitivity to starting values is likely and careful consideration of these values will be required. We examine these issues more carefully in Section 4.

3.3. *Application of principal stratification to estimating modality effects on SFS in the ATM.* To tease out the effect of treatment modality on SFS in the presence of institutionalization, we fit a four strata principal stratification model to our data from the ATM study, where $Z_0$ and $Z_1$ can only take on two values, namely, 0 and 1. Thus, we dichotomized days institutionalized such that $Z_t = 0$ if an individual had 0 days of institutionalization and $Z_t = 1$ otherwise for $t = 0, 1$. We let $t = 1$ and $= 0$ denote residential and outpatient care treatment modalities, respectively. This model allows us to compute treatment effects among adolescents who would not be institutionalized under both treatment and control and among adolescents who would be institutionalized under both treatment and control.

Because SFS has a very skewed distribution with many observed zeros (see Figure 1), we assumed the underlying distribution for $Y$ within each stratum was tobit [Maddala (1983)], e.g., $f(y; \boldsymbol{\theta}_{z_0, z_1, t}) = G(0; \eta^t_{z_0 z_1}, \zeta^2_t)^{1(y \leq 0)} \times g(y; \eta^t_{z_0 z_1}, \zeta^2_t)^{1(y > 0)}$, where $G(y; \eta, \zeta^2)$ and $g(y; \eta, \zeta^2)$ denote the distribution and density functions, resp., for a normal random variable with mean $\eta$ and variance $\zeta^2$. The parameters $\eta^t_{z_0 z_1}$ and $\zeta^2_t$ depend on treatment to allow for treatment effects. As parameterized, $\eta^t_{z_0 z_1}$ and $\zeta^2_t$ are not the mean and variance of the tobit distribution but of the truncated normal which defines the tobit.

Because the ATM is an observational study, there were observable differences in the pretreatment characteristics of youths entering residential and outpatient care. Given these pretreatment differences, any observed differences in treatment group outcomes could result either from differential effectiveness of the treatment modalities, or because of differences in how hard their respective populations are to treat. In order to isolate just the treatment effects of interest, we must compare treatments on equivalent cases. Thus, for the case study, we compare the effectiveness of the two modalities on cases like those in the ATM sample who entered the residential modality. We achieve this comparison by weighting the outpatient sample so that it closely matches the residential sample in terms of the distribution of 86 pretreatment variables expected to be related to substance use and treatment assignment [Morral, McCaffrey and Ridgeway (2004)]. Details on the weighting and comparison of the weighted groups can be found in McCaffrey et al. (2007).

In the remainder of the analysis reported in this example, we compare the unweighted residential sample ($n = 770$) to the weighted outpatient sample ($n = 486$, effective sample size $= 125$), a comparison designed to examine whether the residential modality produces better 12-month outcomes than outpatient care for cases with pretreatment characteristics like those of clients who usually enter residential care. We maximize the weighted likelihood in (3.1) using the EM algorithm. Preliminary results suggested that estimates could be very sensitive to starting values used with the EM algorithm. Consequently, we developed the following approach for selecting starting values.

First, among the residential cases observed to have $Z_1 = 0$, we fit a mixture model of two normal distributions using a simple EM algorithm [Dempster, Laird and Rubin (1977)] to obtain initial estimates of the two means and their corresponding mixing proportions for this group. We labeled the two means $\widehat{\mu}_{0,A}^1$ and $\widehat{\mu}_{0,B}^1$. Under our model, we know that the observed means for residential cases with $Z_1 = 0$ is a mixture of $\mu_{00}^1$ and $\mu_{01}^1$, so $\widehat{\mu}_{0,A}^1$ is an estimate of one of these means and $\widehat{\mu}_{0,B}^1$ is an estimate of the other; however, we do not know which is which. We thus used starting values which allow for both possible mappings in this group.

We repeated this procedure for residential cases with $Z_1 = 1$ and outpatient cases for both $Z_0$ equal to 0 and 1. These steps, in turn, yielded preliminary estimates of the eight mean parameters of the model and their associated mixing proportions. All that remained was to determine how the eight means mapped to the model parameters. There are 16 possible mappings from the preliminary estimates of the simple mixture models to the parameters of the model. For example, one mapping assumes that ($\widehat{\mu}_{0,A}^1$, $\widehat{\mu}_{0,B}^1$, $\widehat{\mu}_{1,A}^1$, $\widehat{\mu}_{1,B}^1$, $\widehat{\mu}_{A,0}^0$, $\widehat{\mu}_{B,0}^0$, $\widehat{\mu}_{A,1}^0$, $\widehat{\mu}_{B,1}^0$) estimates ($\mu_{0,0}^1$, $\mu_{0,1}^1$, $\mu_{1,0}^1$, $\mu_{1,1}^1$, $\mu_{0,0}^0$, $\mu_{1,0}^0$, $\mu_{0,1}^0$, $\mu_{1,1}^0$). The remaining 15 mappings are obtained by switching the mapping of $A$ and $B$ for each observed value of $Z_1$ and $Z_0$ to 0 and 1 exhaustively.

To obtain the final parameter estimates of the model, we ran the EM algorithm to maximize (3.1) separately for each of the 16 possible mappings of the estimates from the simple mixture models and used as final estimates the solution that maximized the log likelihood among all 16 runs. Simulations described below suggested that this procedure can successfully recover the global maximum from the many local maxima.

It is critical that multiple starting values are utilized when applying principal stratification to data. Figure 2 shows the sensitivity of resulting parameter estimates from the 16 different starting values used in our analysis. The results from each set of starting values are plotted as vertical bands and denoted by a number from 1 to 16. Each of the 8 estimated model means is denoted by a row and the resulting estimated values for each mean are plotted for each set of starting values using vertical bars. The first row plots the percentage increase in the negative log-likelihood for each solution compared to the value that minimized it across all sets of starting values (estimate 11 in Figure 2). Each set of starting values leads to a different possible solution, all of which represent a local minimum of the negative log-likelihood in the data. Although the resulting log-likelihood values are very similar (differing by no more than 2 percent), each solution gives very different inferences about the estimated strata means and treatment effects that exist within each stratum. As shown, the solution which gives the minimum negative log likelihood value (denoted by estimate 11 in Figure 2) is distinctly different even from the next best estimate of the model parameters (estimate 6 in Figure 2) which assigns the estimated values of the means for $\mu_{0,0}^1$ and $\mu_{0,1}^1$, $\mu_{1,0}^1$ and $\mu_{1,1}^1$, $\mu_{0,0}^0$ and $\mu_{1,0}^0$, and $\mu_{0,1}^0$ and $\mu_{1,1}^0$ in reverse of how they are assigned in the solution. More generally, it is clear that the alternative solutions tend to find similar values for the various mean values but "flip" the strata labels associated with those labels.

The mixture of tobit models appears to fit the data well, as shown in Figure 3 which plots the histogram of fitted probabilities for each observed condition in our data. Goodness of fit can be noted by the grouping of fitted probabilities at 0 and 1 for the cases of each observed condition. These groupings imply that the majority of our cases have a high probability of falling into one of the two strata of which their observed condition is a mixture. If there had been a more even spread of these values, we would worry about the goodness of fit of the model. Moreover, Table 3 shows that the principal stratification model maintains the marginal means and probabilities of this model.

The estimated treatment effects from the mixture model are shown in Table 4. In contrast to the patterns shown in Table 1, when we control for the confounding effects of institutionalization using principal stratification, residential treatment leads to significantly worse outcomes among adolescents who would experience the same level of institutionalization under both treatment modalities. The effect of residential treatment is larger among adolescents who are not institutionalized under both treatment modalities.

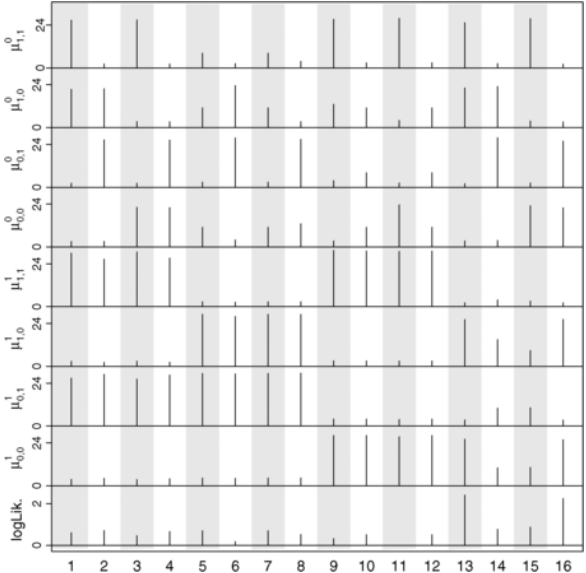

Fig. 2. *The values of the negative log-likelihood and the estimated means for each of the 16 starting values used in the ATM analysis. Each quantity has a separate y-axis and the estimated value of that quantity for each set of starting values is denoted by the heights of the vertical bars.*

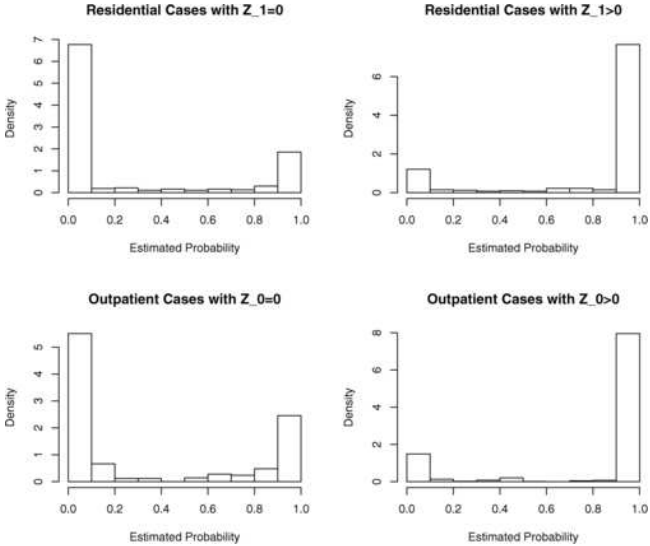

Fig. 3. *Histogram of fitted probabilities for each observed condition.*

The standard errors for treatment effects within each stratum can be adjusted for the effects of clustering (e.g., here adolescents are clustered within treatment sites) by using a Huber–White sandwich estimate of the variance–covariance matrix for the parameters in the likelihood function [Skinner, Holt and Smith (1989)]. When we control for clustering within the ATM data, the treatment effects are no longer significant within the two strata with the same levels of institutionalization under both treatment modalities.

## 4. Evaluation of principal stratification.

Estimating the treatment effects of treatment modality via principal stratification requires identification of latent variables from complex mixture data using mixture proportions. In particular, means for cases from different principal strata are identified from the mixing proportions within observed groups. Given that our exploration of starting values suggested that the identification of strata is potentially weak and that label swapping might be possible, we felt it important to explore the properties of the principal stratification method before interpreting our findings on substance abuse treatment. To our knowledge, there have been very few studies using either real or simulated data sets to explore the convergence of the estimation algorithms, the identification of desired parameters, and the precision of the estimates with varying sample sizes for treatment effects from mixture models like model (3.1).

TABLE 3
*Observed versus predicted marginal means for residential and outpatient care cases*

|  | **Residential care** | | **Outpatient care** | |
|---|---|---|---|---|
|  | **Predicted** | **Observed** | **Predicted** | **Observed** |
| Mean SFS for $Z_{obs} = 0$ | 11.8 | 9.8 | 9.9 | 10.2 |
| Mean SFS for $Z_{obs} = 1$ | 8.1 | 8.2 | 7.8 | 8.1 |
| Proportion institutionalized | 51% | 52% | 43% | 38% |

TABLE 4
*Treatment effect estimates and standard errors comparing residential to outpatient treatment using the four-strata principal stratification model. Significant treatment effects are denoted by * and standard errors are unadjusted for clustering*

|  | $Z_0 = 0$ | $Z_0 > 0$ |
|---|---|---|
| $Z_1 = 0$ | 4.1 (1.7)* | 1.3 (0.9) |
| $Z_1 > 0$ | −1.1 (0.9) | 3.3 (1.8)* |

4.1. *Methods.* To evaluate the method of principal stratification, we began by examining the performance of the four strata model where $Z_0$ and $Z_1$ can only take on two values, namely, 0 and 1. Additionally, we assumed that the underlying distribution of the outcome, $f(y; \boldsymbol{\theta}_{z_0 z_1, t})$, within each strata is normal with mean $\mu^t_{z_0 z_1}$ and variance $\sigma^2_t$, where the means and variances can depend on treatment so that treatment effects can exist. We conducted a simulation study of the properties of parameters estimated by maximizing (3.1) under this assumed parametric model.

First, data was simulated under this model and the impact of sample size, the value of the principal strata probabilities, and the dispersion of the means within each level of $t$ and $Z_t$ on model performance was examined. Second, data was simulated under heavy tailed and skewed distributions and analyzed using the normal model to examine the sensitivity of the normal model to model misspecification. In each case, we maximized the likelihood using the methods described above and compared the resulting estimates to the values used in generating the data.

4.2. *Results.* First, we examined the performance of the estimators for the four strata model under various assumptions about the sample size per treatment arm ($N$), the strata probabilities and the dispersion of the means between strata within the same level of $t$ and $Z_t$. Figures 4, 5 and 6, respectively, plot the estimated means for the control group, the estimated means for the treatment group and the estimated probabilities for each strata versus their true values under the various cases considered. As shown, the method did not begin to perform well until the means within a given level of $t$ and $Z_t$ were at least 1.6 standard deviations away from each other and the sample size within each treatment arm was at least $N = 1000$. See supplementary material for tabulated results [Griffin, McCaffrey and Morral (2008)].

The performance of the estimators was relatively invariant to the true strata probabilities unless the probabilities were uniform, as shown in the third column of Figures 4 and 5. Specifically, we examined three cases for strata probabilities, one in which all four strata had reasonably large probabilities, another in which one strata had a particularly small probability of 0.05, and a third in which the probabilities were equal. Performance was similar in the two cases where the probabilities were not all equal, as shown in the first and second columns of Figures 4 and 5. When the probabilities were equal, the method could not identify the correct mapping of the mixture components to the principal strata values because the model was under-identified. Consequently, the maximum likelihood estimation was unable to recover the model parameters.

We also examined the sensitivity of the four strata model to model misspecification [Griffin, McCaffrey and Morral (2008)]. As expected, the method was sensitive to extremely heavy tailed and skewed distributions. The algorithm appeared to perform well when the data had only moderately heavy tails or moderate skew.

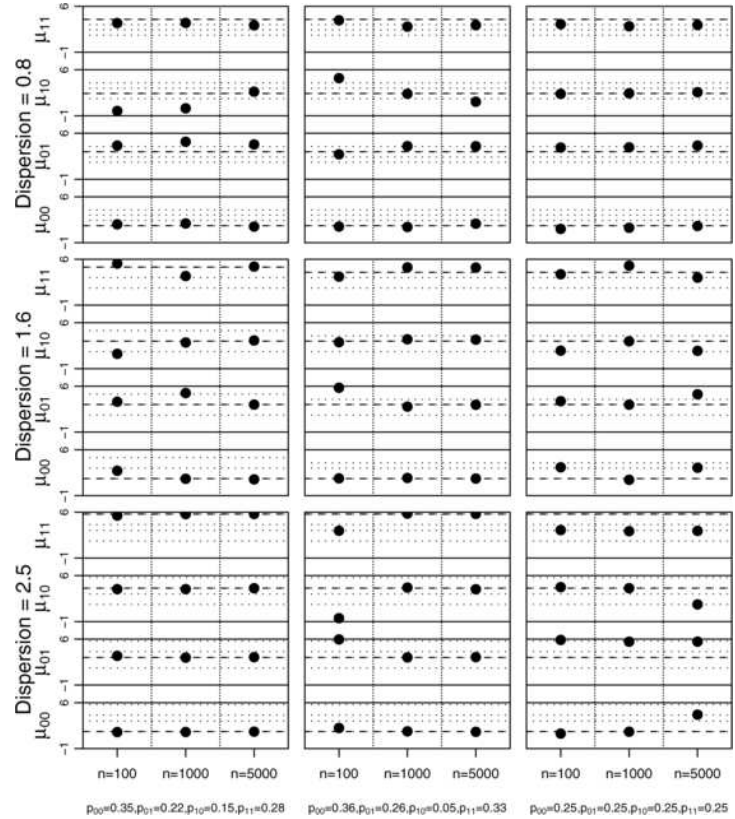

Fig. 4.   *Simulation results. Plot of fitted means (black dots) versus true values (indicated by dashed lines) for the four strata in the control group with means for other strata (indicated as dotted gray lines) for N = 100, 1000 and 5000, different dispersions of the means, and different assumptions about the strata probabilities.*

Unfortunately, our simulation study results suggest that our results for the ATM study must be interpreted with caution. While the dispersion of means within each level of institutionalization in the ATM data is 3.04, the effective sample size in the weighted control group is quite small, only 125, drawing into question the ability of this model to converge to the correct solution if the principal stratification model holds in our data.

Our simulation study analysis was repeated for a nine strata model in which $Z_t$ can take on three values, namely, $0, 1$ and $2$. However, a number of problems were encountered. First, the number of possible sets of starting values (e.g., mappings between the parameters of the full model to estimates from simple mixture models fit to data) increased significantly, leading to $219^2$ possible sets of starting values. Given the computational demands of such a large number of starting values, we did not run the EM algorithm for all possible mappings. Instead we calculated the likelihood at each possible mapping and then tried two techniques for selecting reasonable starting values: (i) choosing the 30 starting values which gave the 30 greatest values of the log-likelihood and (ii) choosing the 30 starting values which represented a spread of the log-likelihood surface. We found that option (i) yielded the best solution (i.e., the solution which gave the highest value of the log-likelihood). However, even with a dispersion of means equal to 2.5 and $N = 5000$ per treatment arm, the best solution from option (i) was unable to recover the model parameters used to simulate the data (results available upon request).

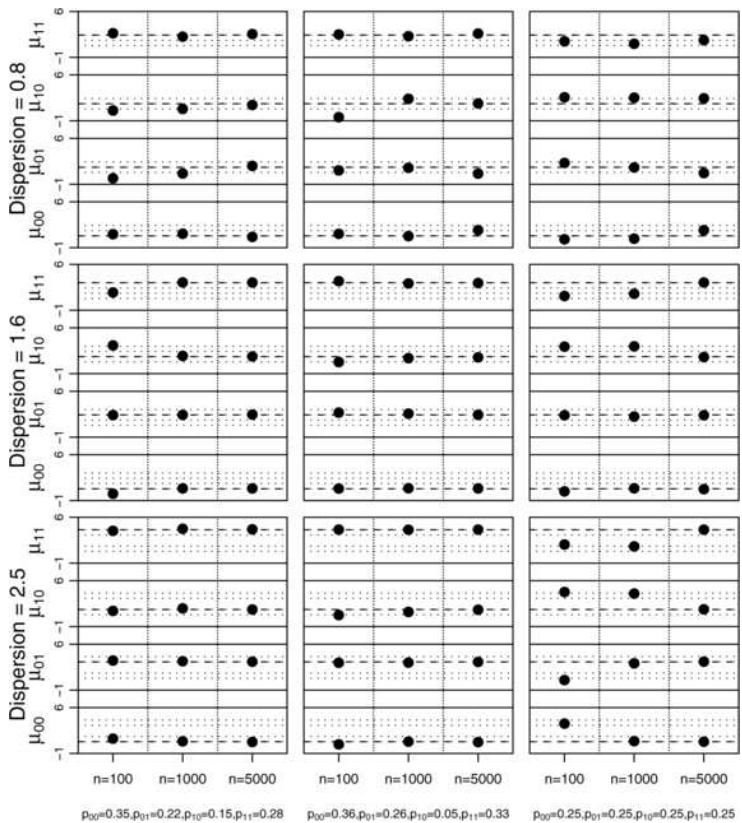

FIG. 5. *Simulation results. Plot of fitted means (black dots) versus true values (indicated by dashed lines) for the four strata in the treatment group with means for other strata (indicated as dotted gray lines) for N = 100, 1000 and 5000, different dispersions of the means, and different assumptions about the strata probabilities.*

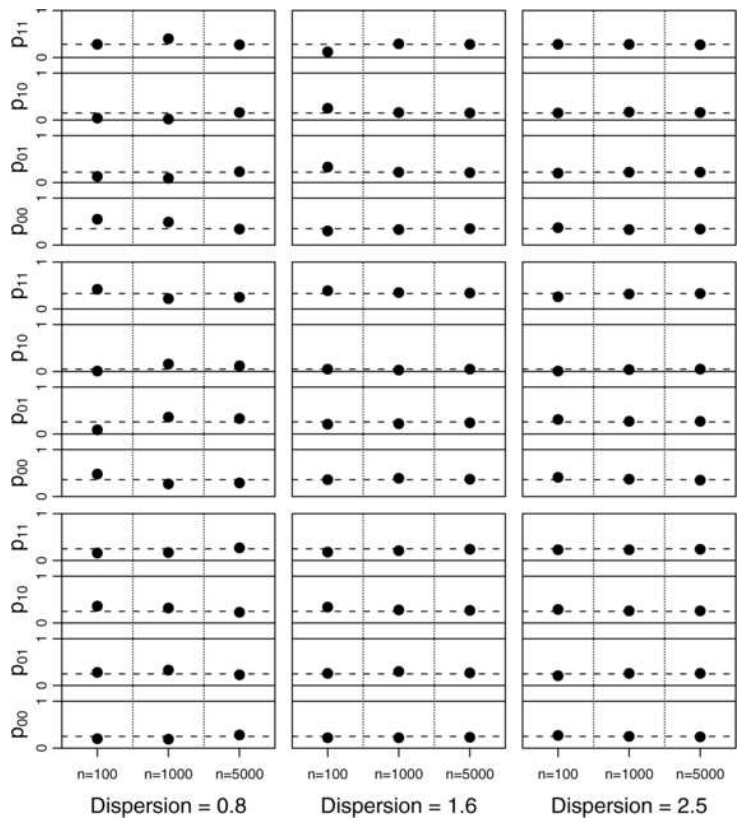

Fig. 6.   *Simulation results. Plot of fitted probabilities (black dots) versus their true values (indicated by dashed lines) for the four strata for N = 100, 1000 and 5000, different dispersions of the means, and different assumptions about the strata probabilities.*

It is likely that adding more parametric assumptions into the nine strata model would help improve the fit of the model. For example, one could consider modeling the means within each stratum and treatment group as a linear function of the value of $Z_1$ and/or $Z_0$. However, as the number of strata increases, more assumptions will be required which may or may not be likely to hold in practice. Inevitably, extending principal stratification to more strata becomes intractable and unidentifiable beyond the simple four strata model.

**5. Discussion.** Institutionalization during the follow-up poses serious challenges to estimating treatment effects on outcomes following substance abuse treatment because it restricts opportunities to use drugs or partake in other problem behaviors (e.g., criminal activity or risky sexual activity). If unaccounted for, it can confound treatment effects and lead to incorrect inferences about the ability of treatment to produce desirable outcomes. Similar problems can occur in many other settings, such as criminal justice and mental health outcomes studies where study participants are also at high risk for institutionalization during the follow-up period.

Because institutionalization occurs post-treatment, treating it as a covariate in analyses can lead to biased results since cases with various levels of institutionalization might differ in terms of their potential outcomes [McCaffrey et al. (2007)]. Most common approaches to the problem require strong assumptions and the results can be very sensitive to which assumptions are made and which methods are used [McCaffrey et al. (2007)]. For example, use of a joint or composite outcome which looks at the effects of treatment on institutionalization and SFS together conflates the effects that treatment has on institutionalization with the effects it has on SFS. Principal stratification allows us to directly model how the treatment effects on SFS may vary within different levels of institutionalization.

Principal stratification provides a framework for developing causal effects on outcomes in the presence of post-treatment institutionalization. The key idea is that causal effects can be obtained by conditioning on cases with equal values for the observed level of institutionalization for the observed treatment status and the unobserved level of institutionalization for the unobserved treatment status. In situations where institutionalization can take on discrete values, the latent principal strata result in finite mixture models and the parameters of interest can potentially be estimated. When applied to adolescent substance abuse treatment data in the ATM study, the method suggests that residential treatment may be less effective for youth who are likely to experience the same level of institutionalization under both conditions (residential and outpatient). However, the effect is strongest among those youth who are unlikely to be institutionalized following treatment— that is, the least problematic users. Alternative analyses [McCaffrey et al. (2007)] found a similar result which improves our confidence in these findings, despite the challenges of the method cited above.

As discussed in the Introduction, stakeholders, like parents, want to know if treatment is effective and for whom it is effective. The principal stratification approach estimates causal effects for youth in different principal strata. Thus, we aim to know if treatment is effective for youth within each principal stratum where institutionalization is constant across treatment conditions. However, the principal strata do not necessarily provide meaningful classification of adolescents to stakeholders. An important follow-up to our analysis is to determine if principal strata membership can be described using measurable and meaningful baseline variables so that stakeholders might have a better idea about which youths will be best served by treatment or different treatment modalities.

Beyond this application, our investigation of principal stratification suggests that the potential of principal stratification may be impossible to realize in many empirical studies. Our study shows that the differences in outcome means among the principal strata must be quite large (1.6 standard deviation units or larger) and that the sample sizes must also be very large ($N \geq 1000$) for estimated effects to correctly identify which means belong to each principal stratum. Unfortunately, such dispersion of means and sample sizes are much greater than those that are likely to arise in many applications. With smaller samples sizes and less dispersed means, group labels are often switched; for instance, the means of the $Z_1 = Z_0 = 0$ stratum might be incorrectly labeled as the means for the $Z_1 = 1$ and $Z_0 = 0$, stratum yielding incorrect inferences about treatment effects. Additionally, the switched mean values may be very plausible so that there would be no indication of misleading results. The limitations of principal stratification must be considered carefully before using the method in all but the most ideal settings.

As the possible values for institutionalization grow, the problem quickly becomes more unmanageable. The number of parameters for the principal strata and the number of mixtures that need to be identified grow rapidly; even finding starting values becomes an extremely challenging task as the possible values for institutionalization grow. Estimation in these contexts will require additional assumptions about the relationship between outcome and institutionalization (e.g., outcomes decline linearly with increased institutionalization) and assumptions about the joint distribution of institutionalization under treatment and control to make the problem more tractable.

Given the challenges of principal stratification when we consider many values for institutionalization, a possible option may be to ignore the outcomes from institutionalized cases and try to estimate an unsuppressed treatment effect restricted to cases that would not be institutionalized under either treatment or control. Our approach to starting values could be used and trustworthy estimates might be obtained in some real world settings. The limitation of this approach would be the inability to generalize the estimates beyond cases that are likely never to be institutionalized. That is, we could not estimate the unsuppressed effect for all cases nor could we determine treatment effects on institutionalized youth.

Using a Bayesian approach might address some of the computational problems with maximizing the likelihood since integrating over a prior helps to reduce sensitivity to the starting values. However, special care would be needed when using Markov chain Monte Carlo methods to sample the posterior to avoid label switching of the mixtures within the observed values of institutionalization [Jasra, Holmes and Stephens (2005)]. Strata label switching will occur because the data provide only very indirect information about the joint distribution of potential institutionalization. Use of an informative prior would be one means of overcoming this limitation of the data via Bayesian analysis. However, this approach is unappealing because analysts are unlikely to be able to make good informed guesses about the joint distribution so that informed priors would be difficult to choose.

Obtaining sufficient information about the joint distribution of the potential institutionalization outcomes is a clear challenge to using the framework of principal stratification to provide useful estimates of treatment effects. One approach to obtaining more information about this distribution would be to collect data at baseline that might be strongly related to the principal strata and use the information as covariates in the model for the strata probabilities. For institutionalization following substance use treatment this information might include criminal activity, detailed information about involvement with the criminal justice system, history with substance abuse treatment, the availability of various types of substance about treatment and sources of payment for the treatment.

Another approach for increasing the information about the principal strata is to jointly model multiple outcomes such as multiple indicators of substance use and criminal activity. Because principal strata are defined by institutionalization and not by other outcomes, principal strata designation should not depend on which outcome is modeled. Combining multiple outcomes can therefore provide more information about the latent principal strata membership and should improve estimation of treatment effects for every outcome. A downside of this approach is the necessity of specifying the joint distribution for multiple outcomes, but the additional modeling might be very beneficial for identifying the principal strata and the accuracy of the resulting treatment effect estimates.

We might consider simply using the principal stratification framework for sensitivity analyses. For example, rather than trying to model the joint distribution of $Z_0$ and $Z_1$, we might specify the joint distribution in terms of the parameters of the marginal distributions of $Z_0$ and $Z_1$ and a parameter specifying their correlation. The value of the correlation parameter could be manipulated to study the sensitivity of the treatment effect estimates to different assumptions about the principal strata. For example, when institutionalization can take on just two values, we could specify the correlation by the interaction from a log-linear model. For continuous institutionalization, we could specify the correlation parameter.

Principal stratification is an important tool for approaching the problem of institutionalization during the follow-up, because it provides a framework for defining estimates of interest and possible methods for estimating them. However, it is also clear that it is not a panacea to the problem because the model parameters are at best weakly identified and it does not extend easily to problems with many possible values for institutionalization. The extensions described above provide useful areas for future research that may significantly improve the application of principal stratification in real world settings.

**Acknowledgment.** The authors are grateful to Rajeev Ramchand for his helpful suggestions on the manuscript.

## SUPPLEMENTARY MATERIAL

**Supplementary tables for "An application of principal stratification to control for institutionalization at follow-up in studies of substance abuse treatment programs"** (DOI: 10.1214/08-AOAS179SUPPA; .pdf). This file contains tabulated results to simulation study of principal stratification method.

**Example data for running principal stratification model in "An application of principal stratification to control for institutionalization at follow-up in studies of substance abuse treatment programs"** (DOI: 10.1214/08-AOAS179SUPPB; .csv). This file contains dataset described in paper.

**Example code for running principal stratification model in "An application of principal stratification to control for institutionalization at follow-up in studies of substance abuse treatment programs"** (DOI: 10.1214/08-AOAS179SUPPC; .txt). This file contains code used to run models in paper.

B. A. Griffin
A. R. Morral
RAND Corporation
1200 South Hayes Street
Arlington, Virginia 22202
USA
E-mail: bethg@rand.org
         morral@rand.org

D. F. McCaffrey
RAND Corporation
4570 5th Avenue
Pittsburgh, Pennsylvania 15213
USA
E-mail: danielm@rand.org